\documentclass[11pt,twoside]{article}

\usepackage{asp2006_m}
\usepackage{epsf}
\usepackage{psfig}
\usepackage{lscape}

\begin{document}
\title{Merging as a key to reforming of disk and AGN triggering in Sy galaxy Mrk334.}   
\author{Aleksandrina Smirnova and Alexei Moiseev}   
\affil{Special Astrophysical Observatory, Nizhnij Arkhyz 369169, Russia
}    

\begin{abstract} 
We have studied the kinematics of the ionized gas and stellar component in Mrk334 using methods of panoramic  (3D) spectroscopy.
The observations were performed at the prime focus of the SAO RAS  6-m telescope   with  the integral-field  spectrograph  MPFS  \citep{mpfs} and
with a scanning  Fabry-Perot interferometer (FPI),  installed on the multimode device SCORPIO \citep{afan}. Based on these data,
the  monochromatic  maps and velocity fields in  different emission lines  of the ionized gas were constructed. The diagnostic
diagrams have been made based on the emission lines ratios.
\end{abstract}

\section{Spectrophotometry and diagnostic diagrams}

\begin{figure}
\plotone{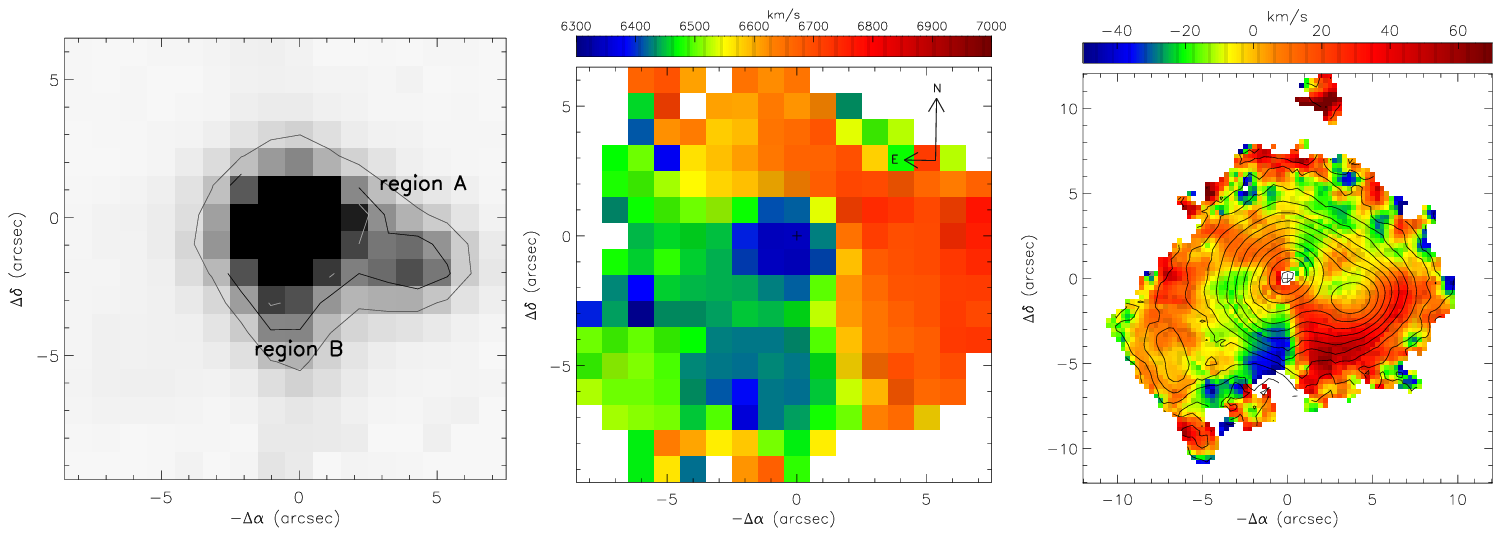}
\caption{ MFPS flux map (left) and velocity field  (middle) in the [OIII]$\lambda$ 5007\AA\, emission line. The large-scale residual
velocity field in the H$\alpha$ line, isophotes in this line are overlapped (right).}
\label{smir:fig1}       
\end{figure}

Mrk334 is a Sy1.8 galaxy, located at the distance of 88 Mpc (the
scale is 430 pc/$''$). This galaxy demonstrates  merging features
as tidal tail  and bright region `A' on the west from the
nucleus. According \citet{gonzadelga97},  the region "A' can be a
second nucleus or a remainder of a satellite galaxy, which was
devoured by Mrk334. On the MPFS maps in different emission lines,
three bright regions can be distinguish: nucleus, regions `A' and
`B' (Fig.~\ref{smir:fig1}). Spectra from these regions differ from
each other and the gas ionization state vary from one region to
another dramatically: on the diagnostic diagrams, the points
belonging to the knot `B' lie in the region of ionization by a
nonthermal radiation. All points that correspond to  the knot `A'
fall into the photoionization region. Surprising, that the nucleus
is located in the HII/LINER boundary i.e., the gas can also be
partially ionized here by both shocks and radiation from young hot
stars, but not by non-thermal source! Also, on the [SII] ratio map
the region with the lowest density is the knot `B', whereas the
density in the knot `A'  is much  higher.

\section{Gas kinematics in the disk of  Mrk334}

Velocity fields corresponding to the most emission lines show almost circular
rotation of ionized gas. Only [OIII]$\lambda 5007$ velocity field is an exception and
reveals  outflow from the nucleus. It becomes apparent in the excess of
blue-shifted velocities in the center of the galaxy. On the large-scale H$\alpha$
residual velocity field (observational velocities minus model of pure circular
rotation) strong non-circular velocity perturbations in the knot `B' are
seen (Fig.~\ref{smir:fig1}, right). The amplitude of the velocity deviations reaches 50\,km\,s$^{-1}$.

\section{What are the regions `A' and `B'?}

On the  deep images obtained on the 6m telescope we find new numerous faint elongated
structures (tidal debris) at different spatial scales. It confirms the merging with a
satellite. This event strongly affects  the disc structure and kinematics  in Mrk334.
Knot `A' can be a region of violent star formation, ionizing diagrams confirm this
conclusion. What is the knot `B' located at the distances of 1-3 kpc from the nucleus?
If  an jet from the active nucleus is placed here, then  we must detect a non-thermal
radio emission. However, radio maps \citep{ulves86} disprove this hypothesis. We suggest
that knot `B' is a region where a satellite debris passing through the disk of the main
galaxy.  This  merging event  can trigger the fueling of  AGN in Mrk 334.

\acknowledgements 
This work was partly supported by the Russian  Foundation for Basic Research (project  06-02-16825) and by the
grant of President of Russian Feredration (MK1310.2007.2)

\end{document}